\documentclass[aps, prl, showpacs, twocolumn]{revtex4-1}

\usepackage{graphics}
\usepackage{dcolumn}
\usepackage{epsfig}
\usepackage{graphicx}
\usepackage{epstopdf}
\usepackage{hyperref}

\usepackage{verbatim}

\begin{document}

\title{Effects of charge doping and constrained magnetization on the electronic structure of an FeSe monolayer}

\author{Timur Bazhirov}
\author{Marvin L. Cohen}
\affiliation{Department of Physics, University of California at Berkeley, Berkeley, California 94720, 
and Materials Sciences Division, Lawrence Berkeley National Laboratory, Berkeley, California 94720}

\date{\today}

%%%%%%%%%%%%%%%%%%%%%%%%%%%%%%%%%%%%%%%%%%%%%%%%%%%%%%%%%%%%%%%%%%%%%%%%%%%%%%%%%%%%%%%%%%%%%%%%%%
%%					ABSTRACT
%%%%%%%%%%%%%%%%%%%%%%%%%%%%%%%%%%%%%%%%%%%%%%%%%%%%%%%%%%%%%%%%%%%%%%%%%%%%%%%%%%%%%%%%%%%%%%%%%%

\begin{abstract}
The electronic structural properties in the presence of constrained magnetization and a charged background are studied for a monolayer of FeSe in non-magnetic, checkerboard-, and striped-antiferromagnetic (AFM) spin configurations. First principles techniques based on the pseudopotential density functional approach and the local spin density approximation are utilized. Our findings show that the experimentally observed shape of the Fermi surface is best described by the checkerboard AFM spin pattern. To explore the underlying pairing mechanism, we study the evolution of the non-magnetic to the AFM-ordered structures under constrained magnetization. We estimate the strength of electronic coupling to magnetic excitations involving an increase in local moment and, separately, a partial moment transfer from one Fe atom to another. We also show that the charge doping in the FeSe can lead to an increase in the density of states at the Fermi level and possibly produce higher superconducting transition temperatures. 
\end{abstract}

\vspace{10pt}

%%%%%%%%%%%%%%%%%%%%%%%%%%%%%%%%%%%%%%%%%%%%%%%%%%%%%%%%%%%%%%%%%%%%%%%%%%%%%%%%%%%%%%%%%%%%%%%%%%
%%					INTRO
%%%%%%%%%%%%%%%%%%%%%%%%%%%%%%%%%%%%%%%%%%%%%%%%%%%%%%%%%%%%%%%%%%%%%%%%%%%%%%%%%%%%%%%%%%%%%%%%%%

\maketitle

Recent experimental advances in molecular beam epitaxy and scanning tunneling microscopy have made it possible to study superconducting monolayer systems, such as FeSe, which is the simplest iron-based superconductor. Studies of FeSe monolayer systems on different substrates show significant sensitivity to interface effects and give signs of the presence of superconductivity above 77 K \cite{fese_ml_sto_first,fese_on_graphene_song}. The latter fact is  especially interesting because bulk samples only show superconducting transition temperatures $T_{\rm c}$ of about 8 K, or 37 K with the application of pressure \cite{hsu_fese, medvedev_fese}. When double-layer graphene is used as a substrate, the superconducting gap seems to increase with the FeSe film thickness, and a monolayer-thick film was found to be non-superconductive \cite{fese_on_graphene_song}. At the same time, when FeSe films are studied on SrTiO$_{3}$, scanning tunneling microscopy reveals the presence of a 20 meV gap for single unit cell thick films, whereas thicker films show no presence of superconductivity \cite{fese_ml_sto_first}.

Another difference between the bulk and monolayer FeSe is the Fermi surface topology: ARPES measurements show distinctive shape where only pockets at the Brillouin zone (BZ) boundaries are present, and no pockets are seen at the center point $\Gamma$ \cite{fese_ml_sto_fs}. The same study shows the presence of large and nearly isotropic superconducting gaps without nodes. Another very recent report by the same research group \cite{fese_2d_2phases_arxiv} describes the presence of two phases with different Fermi surface topology, where a superconducting phase can be obtained from a non-superconducting one through annealing. Both reports give evidences for substantial interface-induced changes in electronic structure. Suppression of superconductivity by twin boundaries interconnected with the Se-atom height with respect to the Fe layer was also found \cite{fese_twin_boundaries}. 

Several theoretical efforts were made to study thin films of FeSe \cite{monolayer_fese_first_princ_liu, fese_sto_dhlee_arxiv}. First-principles study of atomic and electronic structures of one- and two-monolayer thick films on SrTiO$_{3}$ found semiconductor-like behavior and collinear antiferromagnetic order to be present \cite{monolayer_fese_first_princ_liu}. No strong hybridization of the electronic states between the substrate and the film was found. Effects of the interaction between the electronic system and soft phonons of SrTiO$_3$ on FeSe/SrTiO$_3$ interfaces was studied using functional renormalization group calculations \cite{fese_sto_dhlee_arxiv}. The possibility for the soft phonons at the interface to significantly affect the pairing mechanism and increase $T_{\rm c}$ with respect to the bulk sample was proposed.

In the current work we study the electronic structure of an isolated FeSe monolayer in the non-spin-resolved (NM), and for the checkerboard (CH) and the striped (STR) antiferromagnetic (AFM) spin configurations. We utilize a first principle pseudopotential density functional theory and local spin density approximation (LSDA) based approach. Our results show that the shape of the Fermi surface for the doped CH AFM spin configuration is consistent with the experimentally observed one. We study the evolution of the electronic structure from the NM to the AFM cases under magnetization constraints on Fe-atoms. Constrained magnetization-based estimates show that the transfer of magnetic moment between Fe atoms induces a larger bandstructure shift than the case when magnetic moments are increased. The strength of the magnon-induced energy shifts is up to one order of magnitude larger then a typical value for an energy shift caused by phonons. We also study the effect of the introduction of a charge background, and find that it can produce favorable conditions for a higher $T_{\rm c}$ by affecting the number of electrons available for superconducting pairing at the Fermi level.

%%%%%%%%%%%%%%%%%%%%%%%%%%%%%%%%%%%%%%%%%%%%%%%%%%%%%%%%%%%%%%%%%%%%%%%%%%%%%%%%%%%%%%%%%%%%%%%%%%
%%					CALCULATION DETAILS
%%%%%%%%%%%%%%%%%%%%%%%%%%%%%%%%%%%%%%%%%%%%%%%%%%%%%%%%%%%%%%%%%%%%%%%%%%%%%%%%%%%%%%%%%%%%%%%%%%

The electronic properties are calculated using the generalized gradient approximation (GGA) to density functional 
theory (DFT) \cite{dft.method, perdew.burke} within a planewave pseudopotential scheme \cite{cohen.pw,tm.pseudo,mlc_pseudo_physscr}. We use a 40 Ry cutoff value for the kinetic energy of the planewave basis and 600 Ry cutoff - for the electronic density. Crystal structures are relaxed, so that the forces on the atoms are less then 0.5 mRy/$\AA$. For the NM and the CH spin-polarized configuration we use a 4-atom tetragonal unit cell. In the STR spin-polarized case, we choose a unit cell with 8 atoms, and the relaxed structure possesses orthorhombic symmetry as the striped spin arrangement introduces a lattice distortion. The NM, CH and STR phases respectively show: 6.97, 7.10, 7.11 in atomic units (a.u) for the equilibrium lattice constants; 2.57, 2.70 and 2.72 in a.u. for the Se-atom heights; and 0, 2.28 and 2.62 bohr magnetons for the equilibrium magnetic moments. The CH and STR phases are 13 and 20 mRy, correspondingly, lower in total energy then the NM one, when related to the 4-atom cell. The background charge is introduced using a uniform jellium-like approximation. No relaxation is done for the charged configurations. The magnetization constraints are introduced within LSDA through the energy penalty functional, as it is implemented in {\tt Quantum-ESPRESSO} package \cite{pwscf_full}, so that a certain value of magnetization on the individual atoms is preferred.
%%%%%%%%%%%%%%%%%%%%%%%%%%%%%%%%%%%%%%%%%%%%%%%%%%%%%%%%%%%%%%%%%%%%%%%%%%%%%%%%%%%%%%%%%%%%%%%%%%
%%					FIGURE 1
%%%%%%%%%%%%%%%%%%%%%%%%%%%%%%%%%%%%%%%%%%%%%%%%%%%%%%%%%%%%%%%%%%%%%%%%%%%%%%%%%%%%%%%%%%%%%%%%%%

\begin{figure}
\includegraphics[width=\columnwidth]{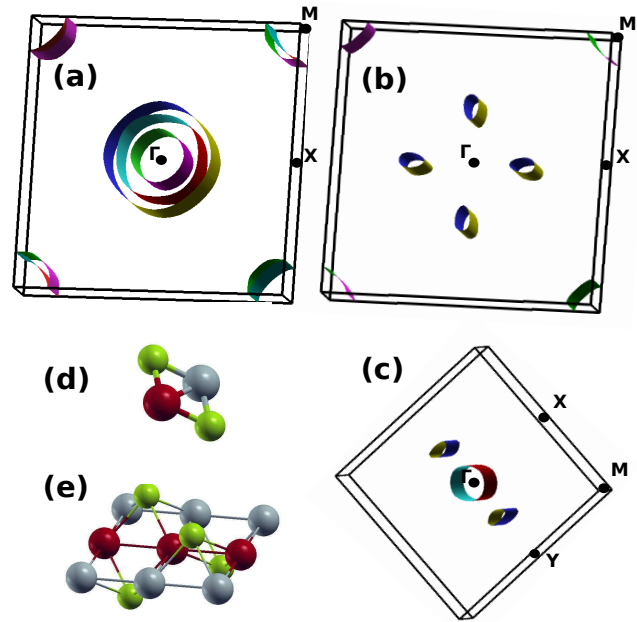}
%width=\columnwidth]{figure2.eps}
\caption{\label{fig1}
(Color online) (a)-(c) Fermi surfaces of an FeSe monolayer for the non-spin-polarized, checkerboard AFM and striped AFM cases respectively. The non-zero length in z-direction is due to the finite size of the unit cell. Brighter colored bands are those crossing the Fermi level. NOTE: the striped spin-configuration has a different BZ (rotated by $\pi/2$ and scaled by $\surd 2$ in $k_x$ and $k_y$-directions). The stripe direction is toward the given X point. (d) - the unit cell with two Fe and two Se atoms, used in our calculations for non-magnetic and checkerboard cases. (e) - the unit cell for the striped iron spin arrangements; the neighboring atoms are also shown. The Fe atoms are red and grey for spin up and down, and Se atoms are shown in green (above and below the Fe-plane).
}
\end{figure}
%%%%%%%%%%%%%%%%%%%%%%%%%%%%%%%%%%%%%%%%%%%%%%%%%%%%%%%%%%%%%%%%%%%%%%%%%%%%%%%%%%%%%%%%%%%%%%%%%%

%%%%%%%%%%%%%%%%%%%%%%%%%%%%%%%%%%%%%%%%%%%%%%%%%%%%%%%%%%%%%%%%%%%%%%%%%%%%%%%%%%%%%%%%%%%%%%%%%%
%%					RESULTS
%%%%%%%%%%%%%%%%%%%%%%%%%%%%%%%%%%%%%%%%%%%%%%%%%%%%%%%%%%%%%%%%%%%%%%%%%%%%%%%%%%%%%%%%%%%%%%%%%%

First, we consider the differences in electronic structure between the bulk and monolayer cases. As can be seen from Fig.~\ref{fig1}(a), the NM Fermi surface for FeSe is very similar to that of the bulk crystal \cite{subediFeSe}. There are three hole pockets at the zone center $\Gamma$ and two electron pockets centered at the corners. Fig.~\ref{fig1}(c) shows the Fermi surface for the STR AFM case, that is similar to the bulk case (for bulk FeSe electronic structure see \cite{own.bulk}, for example). There are two pockets that form the Fermi surface: the hole-pocket is centered at $\Gamma$, and the electron-pocket forms satellites located nearby $\Gamma$ towards the Y direction. The BZ for the STR case is a fraction of those for NM and CH, so we note that the pocket at $\Gamma$ for the STR configuration's reduced BZ would produce pockets at both $\Gamma$ and $M$ for the other two configuration's. The CH case does exhibit a different Fermi surface in the monolayer configuration. Fig.~\ref{fig1}(b) shows only one electronic pocket at $M$ and small hole-like pockets around $\Gamma$ towards $X$. The bandstructure analysis shows that this difference comes mainly from the change in the shape of a "flat" band formed close to the Fermi level $E_F$, below it. In contrast to the bulk case, this band shows a significant reduction in energy width. Analysis of the spatially resolved integrated local density of states in the region of 200 meV below the Fermi level clearly shows the concentration of charge on iron atoms due to the formation of the above-mentioned "flat" band. 

\begin{figure}
\includegraphics[width=\columnwidth]{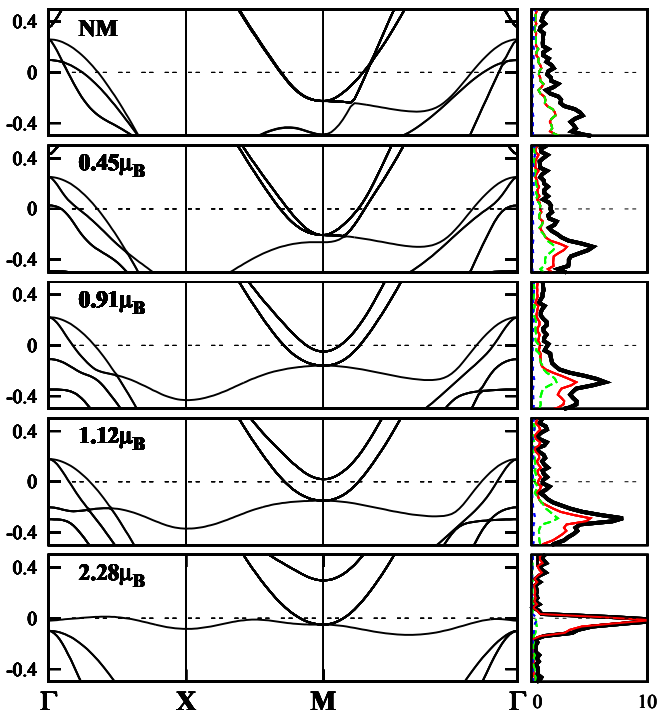}
%width=\columnwidth]{figure2.eps}
\caption{\label{fig2}
(Color online) The evolution of the electronic bandstructure of an FeSe monolayer under constrained magnetization for the checkerboard antiferromagnetic spin-polarized case. From top to bottom the resulting value of the magnetic moment on iron atoms is changing from zero (NM - non-magnetic) to 2.28 $\mu_B$. The units on y-axis are electron-volts, and the Fermi level is at zero. To the right the corresponding projected densities of states (in states/eV) are plotted for all atoms (black solid), Fe-3d-up states (red solid), Fe-3d-down states (green dashed) and Se (blue dot-dashed). 
}
\end{figure}

Secondly, since the experimentally observed value of the Fe magnetic moment for the iron-based superconductors is usually smaller then the first-principle calculations-based theoretical predictions \cite{fese_hjchoi_arxiv,paglione2010}, we introduce constraints on the value of the atomic magnetization to effectively lower the resulting local moment. We then study how the electronic structure evolves from the non-magnetic to spin-ordered case. Fig.~\ref{fig2} shows the resulting evolution for the CH AFM case. It can be seen that when we introduce magnetization, the energy states split around $M$, and as the magnetization increases, a "flat" band starts forming as a composition of the bands both initially present in non-magnetic states around the Fermi level and the lower-lying ones. Similar behavior is seen at the $\Gamma$ point: a state initially at about -0.5eV below the Fermi level rises to form the flat band, and the hole-like bands lower their positions in energy until they fall below $E_F$. As a result, a sharp peak in the density of states is formed right below the Fermi level. Overall, the bandstructure undergoes significant change including both band rearrangements and shifts. We point out here that a similar-looking electronic structure transformation is observed in experiment \cite{fese_2d_2phases_arxiv} via annealing.

%%%%%%%%%%%%%%%%%%%%%%%%%%%%%%%%%%%%%%%%%%%%%%%%%%%%%%%%%%%%%%%%%%%%%%%%%%%%%%%%%%%%%%%%%%%%%%%%%%
%%					FIGURE 2
%%%%%%%%%%%%%%%%%%%%%%%%%%%%%%%%%%%%%%%%%%%%%%%%%%%%%%%%%%%%%%%%%%%%%%%%%%%%%%%%%%%%%%%%%%%%%%%%%%

%%%%%%%%%%%%%%%%%%%%%%%%%%%%%%%%%%%%%%%%%%%%%%%%%%%%%%%%%%%%%%%%%%%%%%%%%%%%%%%%%%%%%%%%%%%%%%%%%%

%%%%%%%%%%%%%%%%%%%%%%%%%%%%%%%%%%%%%%%%%%%%%%%%%%%%%%%%%%%%%%%%%%%%%%%%%%%%%%%%%%%%%%%%%%%%%%%%%%
%%					FIGURE 3
%%%%%%%%%%%%%%%%%%%%%%%%%%%%%%%%%%%%%%%%%%%%%%%%%%%%%%%%%%%%%%%%%%%%%%%%%%%%%%%%%%%%%%%%%%%%%%%%%%

\begin{figure}
\includegraphics[width=\columnwidth]{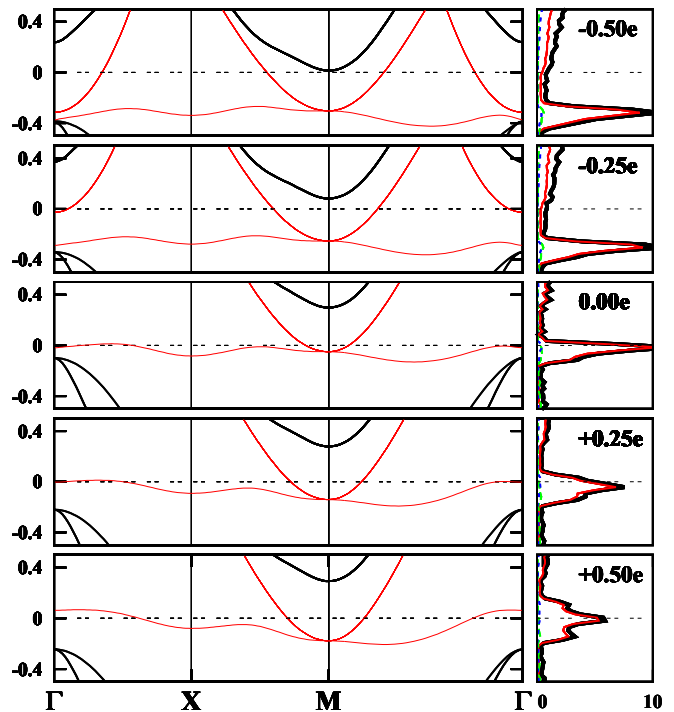}
%width=\columnwidth]{figure2.eps}
\caption{\label{fig3}
(Color online) The electronic bandstructures of the FeSe monolayer with different levels of charge doping in the checkerboard antiferromagnetic spin-polarized case (with 2.28 $\mu_B$ and no constraints). From top to bottom the value of excessive background charge is changing from -0.5e to +0.5e, as shown. The units on the y-axis are electron-volts, and the Fermi level is at zero. To the right the corresponding projected densities of states (in states/eV) are plotted for all atoms (black solid), Fe-3d-up states (red solid), Fe-3d-down states (green dashed) and Se (blue dot-dashed). Red-colored (brighter-colored) bands are the ones crossing the Fermi level.
}
\end{figure}

%%%%%%%%%%%%%%%%%%%%%%%%%%%%%%%%%%%%%%%%%%%%%%%%%%%%%%%%%%%%%%%%%%%%%%%%%%%%%%%%%%%%%%%%%%%%%%%%%%

Since one of the possible effects of the interface could be a charge exchange, it is interesting to analyze how the electronic structure is affected by doping. To do this, we introduce a uniform jellium charge backgroung. Fig.~\ref{fig3} shows the results for the CH AFM case (with 2.28 $\mu_B$ and no constraints), where two particularly interesting things can be noted: first, the width of the "flat" band formed below $E_F$ is very sensitive to positive doping, and hole-like states appear at the Fermi level at various point of the BZ; secondly, an electron-like state falls below the Fermi level at $\Gamma$ point for negative doping, forming a second electron-like pocket of the Fermi surface. The latter configuration has a Fermi surface that is similar to what was obtained for a bulk intercalated FeSe-compound in \cite{tl_rb_fese_ellike_fs}. The density of states plots given in Fig.~\ref{fig3} exhibit a shape, that favors an increase at $E_F$ for both positive and negative doping. In the former case the increase is induced by moving the Fermi level into the flat-band-originated peak, in the latter case the increase is due to the appearance of the above-mentioned electron-like band at $\Gamma$. For the NM and the STR phases our calculations show a rigid band shift of $E_F$ for the positive charge doping. The hole-like bands at $\Gamma$ exhibit higher sensitivity in their relative position with respect to the Fermi level, shifting upward in energy. For the negative charge doping we see new electron-like states appear at the Fermi level around $\Gamma$ for -0.25e in the non-magnetic case, and for -0.75e doping in the striped case. The density of states at the Fermi level shows a relative increase for both doping directions in both NM and STR phases.

%%%%%%%%%%%%%%%%%%%%%%%%%%%%%%%%%%%%%%%%%%%%%%%%%%%%%%%%%%%%%%%%%%%%%%%%%%%%%%%%%%%%%%%%%%%%%%%%%%
%%					DISCUSSION
%%%%%%%%%%%%%%%%%%%%%%%%%%%%%%%%%%%%%%%%%%%%%%%%%%%%%%%%%%%%%%%%%%%%%%%%%%%%%%%%%%%%%%%%%%%%%%%%%%

There appears to be a general belief that the non-spin-resolved structure studies using the local-density approximation provide Fermi surface shapes that are in agreement with experiment at least for 1111 and 122 compounds  %\cite{singhdu_lafeaso,lebeague_lafepo,lafeaso_fs_arpes,lafepo_arpes_fs_2008},
\cite{fese_hjchoi_arxiv, paglione2010}. Nevertheless, some properties such as equilibrium lattice constants, for example, are better described within LSDA \cite{cao_kfese_fs_dft, yildirim_lsda_lattconst_prl_2009}. On the basis of our findings we suggest here that a specific AFM spin orientation might be induced in the FeSe layer by the substrate. There have been previous studies that reported that SrTiO$_3$ substrate can induce AFM order at the interface in thin films \cite{stio_afm_domains_films, stio_coexistense_li_2011}. The existence of only one $M$-point centered Fermi surface pocket was addressed with respect to the superconducting pairing in \cite{kfese_pairing_2011_dhlee} and might also mean that the currently widely accepted $s_\pm$ model \cite{mazin_s_pm, mazin_kfese_2011_symmetry} needs to be adjusted.

To explore the possible pairing mechanisms, we have studied the bandstructure evolution with magnetization constraints, and analyze the bandshifts introduced by particular magnetic excitations. This approach is analogous to a frozen phonon approach to evaluate electron-phonon coupling. For example, from Fig.~\ref{fig2} we extract the information for the relative position of the bottom of the electron-like band at $M$ with respect to $E_F$ as a function of the Fe magnetic moment. In the unfolded non-magnetic BZ of iron lattice only (similar to used in \cite{kfese_pairing_2011_dhlee}) the introduction of the CH spin orientation corresponds to the magnon $q$-vector of ($\pi$,$\pi$). Similarly for the evolution of the NM to the STR configuration we analyze the relative position of the hole-like band maxima at $\Gamma$. This excitation corresponds to a ($\pi$, 0) magnon in the unfolded BZ. Both ($\pi$,$\pi$) and ($\pi$, 0) magnons involve an increase of local Fe moments in a corresponding spin arrangements with respect to the non-magnetic configuration. We also apply a small-magnitude magnetization shift from ($\mu_0$, $-\mu_0$) to ($\mu_0 + \delta\mu$, $-\mu_0 + \delta\mu$), corresponding to an effective transfer of local moment between the Fe atoms. %near-zero $q$-vector excitations.
Our analysis shows that the energy shifts for all three cases do not strongly depend on the value of the iron-magnetic moment. The relative strengths of the shifts are estimated as follows:

\begin{equation}
\label{Bandshifts}
(\frac{\partial E_M}{\partial \mu})_{(\pi,\pi)}:(\frac{\partial E_{\Gamma}}{\partial \mu})_{(\pi, 0)}:(\frac{\partial E}{\partial \mu})_{(trans)} = 1 : 1.5 : 6
\end{equation}
In addition, we analyze the $(\partial \mu / \partial a)$ derivative, where $a$ is the in-plane lattice constant, to calculate $(\partial E/ \partial a)_{mag}$ = $(\partial E/ 
\partial \mu)(\partial \mu/ \partial a)$. To accomplish this we obtain the relaxed structures for each magnetization value considered. The estimates give $(\partial E/ \partial V)_{(\pi,\pi)}$ $\approx$ 1 eV/$\AA$, which is of the same order as a characteristic value for a frozen phonon, if we consider $(\partial E)_{ph}$ $\approx$ 15 meV, $a$ $\approx$ 3.75$\AA$, which are typical for FeSe, and a $(\partial a /a)_{ph}$ $\approx$ 0.005. So we can conclude that the local moment transfer-involving magnetic excitation are coupled to electronic states stronger than the excitations that involve magnetization increase only. The relative coupling value for the latter is of the same order as for the phonons.

As seen in the current work, the increase in the superconducting gap value may originate from the charge-transfer-induced increase in the density of states at the Fermi level for the FeSe-layer, and the charge transfer itself can originate from the interaction with the SrTiO$_3$ substrate, or from chemical doping in the case of alkali doped FeSe materials. Such an increase would mean that there are more electronic states that can take part in the scattering process forming Cooper pairs, thus increasing the gap between the superconducting and normal states. Controlling the charge carriers density in thin films by gating has long been an actively researched field for graphene. The results here suggest that similar approaches can be beneficial for studying superconductivity in FeSe and possibly in other iron-based superconductors. Our results also suggest that the superconducting gap in an FeSe monolayer can be enhanced with gating.

%%%%%%%%%%%%%%%%%%%%%%%%%%%%%%%%%%%%%%%%%%%%%%%%%%%%%%%%%%%%%%%%%%%%%%%%%%%%%%%%%%%%%%%%%%%%%%%%%%
%%					CONCLUSION
%%%%%%%%%%%%%%%%%%%%%%%%%%%%%%%%%%%%%%%%%%%%%%%%%%%%%%%%%%%%%%%%%%%%%%%%%%%%%%%%%%%%%%%%%%%%%%%%%%

In conclusion, we have studied the electronic properties of an FeSe monolayer in different spin configurations. Our results show that the checkerboard antiferromagnetic spin pattern yields a Fermi surface that resembles the experimental results for FeSe on STiO$_3$. This suggests that a similar antiferromagnetic spin pattern is induced in the monolayer under experimental conditions. A study of the evolution of the non-magnetic electronic structure into the checkerboard and the striped antiferromagnetic configurations under constrained magnetization reveals that the magnetic excitations involving local moment transfer are coupled to electronic states stronger then the ones involving an equal change in magnetization. By simulating the substrate-induced interface effects we show that the presence of a uniform positive background charge in the FeSe-layer can lead to an increase in the density of states at the Fermi level producing favorable conditions for a higher superconducting transition temperature. Our findings propose that gated devices similar to the ones used now for graphene may be able to reveal new interesting results in FeSe and other two-dimensional superconductors.

%%%%%%%%%%%%%%%%%%%%%%%%%%%%%%%%%%%%%%%%%%%%%%%%%%%%%%%%%%%%%%%%%%%%%%%%%%%%%%%%%%%%%%%%%%%%%%%%%%
%%					ACKNOWLEDGEMENT
%%%%%%%%%%%%%%%%%%%%%%%%%%%%%%%%%%%%%%%%%%%%%%%%%%%%%%%%%%%%%%%%%%%%%%%%%%%%%%%%%%%%%%%%%%%%%%%%%%

This work was supported by National Science Foundation Grant No. DMR07-05941 and by the Director, Office of Science, Office of Basic Energy Sciences, Division of Materials Sciences and Engineering Division, U.S. Department of
Energy under Contract No. DE- AC02-05CH11231.  Computational resources have been provided by LBNL. We want to thank Dung-Hai Lee, Steven G. Louie and Kevin Chan for fruitful discussions.

\bibliography{references_FeSe}
\bibliographystyle{apsrev4-1}

\end{document}